# Metal Oxidation Kinetics and the Transition from Thin to Thick Films

Zhijie Xu[1,a)], Kevin M. Rosso[2] and Stephen Bruemmer[3]

1. Computational Mathematics Group, Fundamental and Computational Sciences Directorate, Pacific Northwest National Laboratory, Richland, WA 99352, USA

2. Chemical and Materials Sciences Division, Fundamental and Computational Sciences Directorate, Pacific Northwest National Laboratory, Richland, WA 99352, USA

3. Energy Science & Technology Directorate, Pacific Northwest National Laboratory, Richland, WA 99352, USA

We report an investigation of growth kinetics and transition from thin to thick films during metal oxidation. In the thin film limit (< 20 nm), Cabrera and Mott's theory is usually adopted by explicitly considering ionic drift through the oxide in response to electric fields, where the growth kinetics follow an inverse logarithmic law $\log(dl/dt) \propto 1/l$. It is generally accepted that Wagner's theory, involving self-diffusion, is valid only in the limit of thick film regime (>1μm) and leads to parabolic growth kinetics $dl/dt \propto 1/l$, where $l$ is the oxide film thickness. Theory presented here unifies the two models and provides a complete description of oxidation including the transition from thin to thick film. The range of validity of Cabrera and Mott's theory and Wagner's theory can be well defined in terms of the Debye-Hückel screening length. The transition from drift-dominated ionic transport for thin film to diffusion-dominated transport for thick film is found to strictly follow the direct logarithmic law $\log(dl/dt) \propto -l$ that is frequently observed in many experiments.

---

a) Electronic mail: zhijie.xu@pnnl.gov

## I. Introduction

Oxidation or corrosion is often a performance-limiting concern for many materials used in hostile environments. For example, the "corrosion" process taking place on the metal surface often causes undesirable loss of mass that is similar to the precipitation/dissolution process.[1, 2] A layer of oxide film (scale) is often formed under certain conditions and provides the underlying metal protection against further corrosion provided that the scale does not crack or spall.

The material oxidation kinetics has been an interesting topic and extensively studied both theoretically and experimentally in the last decade. In the classical vision of oxide film growth,[3] metal oxidation involves both electronic and ionic transport across the oxide film in order to ionize the metal at the oxide-metal interface and oxidizing agent (for example, oxygen) at the gas-oxide interface, as shown in Fig. 1. It depends on the detailed oxide structure whether the transport of metal ions dominates over that of oxygen ions, or whether the electronic current is mostly carried by electrons or by holes. For our purposes of developing a general theory, the essential aspects include (*c.f.*, Fig. 1):

i. Electrons are assumed to be more mobile than metal ions $M^+$ and are transported fast enough to ionize oxygen molecules at the gas-oxide interface. An electrical field $E$ is accordingly developed across the film that can speed up the metal ion transport.

ii. Metal ions $M^+$ are assumed to be more mobile in the oxide than oxygen ions $O^{2-}$ so that new oxide is always formed at the gas-oxide interface (as opposed to the oxide-metal interface). Recently studies on the oxide film grown on zirconium conclude the transport of oxygen to the oxide-metal interface is dominating at the early stage of oxidation.[4] This situation, though different from our assumption where metal ions are more mobile, still can be studied by following the same methodology.

iii. Oxidation reaction at the gas-oxide interface is fast compared to the transport of metal ions across the oxide film. Continuous production of oxide from the reaction leads to a moving gas-oxide interface at a velocity of $V_s = dl/dt$.

Fundamental treatments of metal oxidation can be dated back to the early 1900s.[5-7] They established the classical parabolic growth kinetics

$$l^2 = k_p t \text{ or } dl/dt = k_p/(2l), \tag{1}$$

where $l$ is the film thickness, $t$ is time of oxidation, and $k_p$ is the parabolic rate constant. The results are summarized in the review article by Cabrera and Mott.[8] Wagner's theory[7] considers self-diffusion of ions across the oxide film is the slowest relevant process and therefore the rate-limiting process. In this theory, the parabolic rate constant $k_p$ can be related to some measurable transport coefficients of the oxide film. This simple model, though not sufficient to predict all practical applications, is very useful for understanding the most important features of oxidation and gain essential knowledge for more complex systems.

The electrical field developed across the oxide during growth is another important feature in the sense that it can facilitate the transport of ions.[3] It was regarded as arising from ambipolar diffusion of the positive and/or negative ions and electrons in the oxide. It is generally accepted that Wagner's theory and parabolic growth kinetics are valid for thick films roughly greater than 1 μm, where $E$ is small enough and the condition for Nernst-Einstein relation ($qEa \ll kT$) is still valid.[3] In this condition, $q$ is the charge of ions, $a$ is the elementary ionic jump distance on the same order as lattice parameter, $k$ is the Boltzmann constant, and $T$ is the temperature. For thin films, the strong electrical field $E$ and attendant breakdown of the Nernst-Einstein relation invalidate Wagner's theory. Cabrera and Mott further developed the theory for thin film growth where the oxidation rate was only controlled by the ionic jump process in the presence of the electric field.[8] Through the calculation of the kinetic barrier associated with the ionic jump, they were able to show that the oxidation rate decreases exponentially as film thickness increases, namely the growth kinetics follow an inverse logarithmic law of

$$\log(dl/dt) = B_1 + B_2/l. \tag{2}$$

A typical plot of growth rate $V_s$ various with film length calculated from both theories is presented in Fig. 2,[3] with the dashed line representing the unknown transition from thin film to thick film growth that will be addressed by this study.

Despite the success of both theories on growth kinetics for metal oxidation, considerable experiments do support a direct logarithmic rate law of

$$\log(dl/dt) = B_1 - B_2 l \text{ or } l = \frac{1}{B_2}\log\left(e^{B_1}B_2 t + 1\right) \tag{3}$$

where $B_1$ and $B_2$ in Eqs. (2) and (3) are constants. Some examples are nickel oxidation at 200°C by Scheuble,[9] the oxidation of aluminum at 25°C by Hart,[10] the oxidation of iron at 25°C by Kruger and Yolken,[11] the oxidation of single crystal copper,[12] and most recently, the low temperature oxide growth on Al single crystals based on molecular dynamics simulation.[13] The direct logarithmic rate law is even more frequently observed and documented than the inverse logarithmic relationship in the literature.[14] In addition, some experiments observed a decrease in the parabolic rate constant $k_p$ with increasing film thickness across all temperatures.[15] Such findings also implied that the growth kinetics cannot be described by the standard Wagner's theory without considering the transition behavior.

This paper presents a generalized oxidation model that recovers Cabrera and Mott's theory at the thin film limit involving drift-dominated ionic transport and Wagner's theory at the thick film limit where diffusive ionic transport is the dominant mechanism. The transition between the two is shown to strictly follow a direct logarithmic law, consistent with the aforementioned experimental and molecular dynamics studies. Therefore, depending on the stage of oxidation, three growth kinetics regimes are unified into a single model with the oxidation starting with a drift-dominated regime (described by the inverse logarithmic law), followed by a transitional regime, and ending with the diffusion-dominated regime (described by a parabolic rate law). For the first time, two different oxidation kinetics describing the oxidation growth rate at the initial stage for thin film and the final stage for thick film were unified into a single complete description, where the transition from thin to thick film can be clearly identified.

## II. Generalized kinetic model for metal oxidation

The generalized kinetic model should provide governing equations for the moving gas-oxide interface during metal oxidation. The simplest oxidation model includes ionic transport in the oxide film from the oxide-metal interface (denoted by $\Gamma_2$) to the gas-oxide interface (denoted by $\Gamma_1$), as shown in Figure 1. The oxide growth kinetics during oxidation is a result of sustaining both ionic transport within the oxide and oxidation reaction (for example, $M^{2+} + O^{2-} \rightarrow MO$) at the gas-oxide interface.

Since the Nernst-Einstein relationship is no longer valid for films less than 20 nm, we will start from the ion jump in presence of strong electrical field that is created by electron transfer from the metal to the gas-oxide interface to ionize the oxygen molecules. Suppose the ion jump rate $\gamma$ follows the standard relationship

$$\gamma = \nu \cdot \exp\left(\frac{-W}{kT}\right), \tag{4}$$

where $\nu$ is the vibrational frequency, $T$ is the temperature, $W$ is the energy barrier for a jump, and $k$ is the Boltzmann constant. For diffusion in one dimension (the case in our study), the ion diffusion coefficient $D_A$ without an electrical field can be expressed as

$$D_A = \frac{\Gamma a^2}{2} = \frac{a^2 \nu}{2} \exp\left(\frac{-W}{kT}\right), \tag{5}$$

where $a$ is the ionic elementary jump distance. Considering the one-dimensional ionic jump along both forward and backward directions in the presence of an electrical field (as shown in Fig. 3), the ion drift velocity $V$ can be written as

$$V = \frac{a\Gamma}{2}\exp\left(-\frac{W}{kT}\right)\left[\exp\left(\frac{qaE}{2kT}\right) - \exp\left(-\frac{qaE}{2kT}\right)\right] = a\nu \exp\left(-\frac{W}{kT}\right)\sinh\left(\frac{qaE}{2kT}\right), \tag{6}$$

where $q$ is the ion charge, $E = \partial\varphi/\partial x$ is the electrical field strength, and $\varphi$ is the electrical potential field. The ion drift velocity can be expressed in terms of the ion diffusion by

$$V = \frac{2D_A}{a}\sinh\left(\frac{qaE}{2kT}\right). \tag{7}$$

It is clear that the drift velocity is zero without electrical field ($E = 0$) and recovers the Nernst-Einstein relationship

$$V \approx \mu E = \frac{qD_A}{kT}E \tag{8}$$

in the presence of a weak electrical field ($qEa \ll kT$), where $\mu$ is the ion mobility. Equation (7) describes the ion drift in terms of the ion diffusion in a more general form than Eq. (8). The transport equation for metal ions within the oxide can be written as

$$\frac{\partial C_A}{\partial t} + V \frac{\partial C_A}{\partial x} = D_A \frac{\partial^2 C_A}{\partial x^2}, \tag{9}$$

where $C_A$ is the concentration of metal ions $M^+$. Considering the flux of metal ions into the gas-oxide interface ($\Gamma_1$), the mass balance at $\Gamma_1$, the moving velocity (oxide growth rate) of interface $\Gamma_1$ and the ion concentration at the interface is written as

$$V_s = -\frac{D_A}{\rho} \frac{\partial C_A}{\partial x}\bigg|_1^- \quad \text{and} \quad C_A\big|_1^- = 0 \qquad \text{on } \Gamma_1, \tag{10}$$

where $\rho$ is the molar density of oxide product with a unit of mol/m$^3$. The zero ion concentration at the interface is the result of the assumed fast oxidation reaction (i.e., fast ion consumption) compared to ionic transport. At the metal-oxide interface $\Gamma_2$, we have a fixed ion concentration

$$C_A\big|_2^+ = C_\infty \qquad \text{on } \Gamma_2. \tag{11}$$

$(\partial C_A/\partial x)\big|_1^-$ in Eq. (10) is the ion concentration gradient at interface $\Gamma_1$ with $|^-$ indicating the magnitude at the oxide side of the interface. $C_A\big|_2^+$ in Eq. (11) is the ion concentration at the interface $\Gamma_2$ with $|^+$ indicating the magnitude at the oxide side of the interface. Equation (10) gives the interface moving velocity that can be derived from local mass conservation. Equations (9), (10), and (11) provide a complete mathematical model for oxidation kinetics that can be analytically solved.

First, these equations are rewritten in dimensionless form by introducing the unit of length $a$, unit of time $a^2/D_A$, unit of velocity $U = D_A/a$, and a dimensionless péclet number $p_e$,

$$\frac{\partial c_A}{\partial t} + p_e \frac{\partial c_A}{\partial x} = \frac{\partial^2 c_A}{\partial x^2}, \tag{12}$$

At the interface,

$$v_s = -\left.\frac{\partial c_A}{\partial x}\right|_1^- \quad \text{and} \quad \left.c_A\right|_1^- = 0 \text{ on } \Gamma_1, \tag{13}$$

and

$$\left.c_A\right|_2^+ = c_\infty \quad \text{on } \Gamma_2. \tag{14}$$

Ion concentration is normalized by $c_A = C_A/\rho$, the molar density of the oxide product. The péclet number $p_e$ is expressed as

$$p_e = 2\sinh\left(\frac{qaE}{2kT}\right). \tag{15}$$

It is not trivial to find the variation of electric field $E$ with the normalized film thickness $L = l/a$. Even at equilibrium, the electric field will not be uniform in general. Separation of charged particles results in surface charges at both interfaces and space charges of opposite polarity distributed in the oxide film. At equilibrium the Debye-Hückel equation can be used to estimate the electrical potential field $\varphi$

$$\frac{\partial^2 \varphi}{\partial x^2} = \frac{\varphi}{L_D^2} \tag{16}$$

with the boundary conditions $\varphi(x=L) = -\varphi(x=0) = \varphi_0/2$. $L_D$ is the normalized Debye-Hückel screening length, and in principle can be related to the temperature $T$ and total charge concentration. A solution of $\varphi$ to Equation (16) can be obtained as

$$\varphi(x) = \frac{\varphi_0 \sinh\left[(x-L/2)/L_D\right]}{2\sinh\left[L/(2L_D)\right]}, \tag{17}$$

and therefore the electric field is

$$E = \frac{\partial \varphi}{\partial x} = \frac{\varphi_0 \cosh\left[(x-L/2)/L_D\right]}{2L_D \sinh\left[L/(2L_D)\right]}. \tag{18}$$

A plot of $\varphi$ variation with $x$ for various oxide thickness $L$ is presented in Fig. 4. For thin film ($L \ll L_D$), $\varphi$ can be approximated by

$$\varphi(x) \approx \varphi_0 (x/L - 1/2) \text{ for } L \ll L_D \tag{19}$$

and leads to a uniform electric field $E$ within the oxide. For thick film ($L \gg L_D$), $\varphi$ can be approximated by

$$\varphi(x) = \frac{\varphi_0}{2}\exp(-x/L_D)\left\{1 - \exp\left[(2x-L)/L_D\right]\right\} \text{ for } L \gg L_D \tag{20}$$

and leads to an almost zero electric field everywhere except the narrow region nearby two interfaces. It is clear that the assumption used in Wagner's theory, i.e. the electrical neutrality in most part of film, is valid for $L \gg L_D$

We now can derive the expression of the péclet number $p_e$ from Eq. (15) using the electric field at $x = L/2$, where

$$E\left(x = \frac{L}{2}\right) = \frac{\varphi_0}{2L_D \sinh\left[L/(2L_D)\right]}, \tag{21}$$

and

$$p_e = 2\sinh\left[\frac{\alpha}{4L_D \sinh(L/2L_D)}\right].  \tag{22}$$

The dimensionless number $\alpha = q\varphi_0/kT$ represents the ratio of the electric potential energy compared to the thermal energy. $\alpha$ is on the order of 10-100 with $\varphi_0$ typically being a few volts.

In principle, Equations (12)-(14) together with the expression of péclet number $p_e$ from Eq. (22) can be solved using a level set method,[16] a phase-field approach,[1] or any other interface tracking methods. In order to solve it analytically and provide more insights, we will follow our method (Reduced-boundary-function method[17]) in a previous study[18] for thermal oxidation.

### III. Solutions to the general kinetic model for metal oxidation

In order to solve Eqs. (12)-(14), we first introduce the following relationships between the interface values and the interface moving velocity through a straightforward differential analysis, as shown in Fig. 5:

$$\left.\frac{\partial c_A}{\partial t}\right|_1^- = \left.\frac{\partial c_A|_1^-}{\partial t}\right. - \left.\frac{\partial c_A}{\partial x}\right|_1^- \cdot v_s,  \tag{23}$$

$$\left.\frac{\partial(\partial c_A/\partial x)}{\partial t}\right|_1^- = \left.\frac{\partial(\partial c_A/\partial x)|_1^-}{\partial t}\right. - \left.\frac{\partial^2 c_A}{\partial x^2}\right|_1^- \cdot v_s.  \tag{24}$$

Similarly, other higher order derivatives at interface $\Gamma_1$ can be obtained in the same fashion and are written as,

$$\left.\frac{\partial\left(\partial^{n}c_{A}/\partial x^{n}\right)}{\partial t}\right|_{1}^{-} = \left.\frac{\partial\left(\partial^{n}c_{A}/\partial x^{n}\right)}{\partial t}\right|_{1} - \left.\frac{\partial^{n+1}c_{A}}{\partial x^{n+1}}\right|_{1}^{-} \cdot v_{s}. \tag{25}$$

Using Eq. (12), we can write the derivatives at interface $\Gamma_1$ as,

$$\left.\frac{\partial^{n+2}c_{A}}{\partial x^{n+2}}\right|_{1}^{-} = \left.\frac{\partial^{n}\left(\partial c_{A}/\partial t\right)}{\partial x^{n}}\right|_{1}^{-} + p_{e}\left.\frac{\partial^{n+1}c_{A}}{\partial x^{n+1}}\right|_{1}^{-} = \left.\frac{\partial\left(\partial^{n}c_{A}/\partial x^{n}\right)}{\partial t}\right|_{1}^{-} + p_{e}\left.\frac{\partial^{n+1}c_{A}}{\partial x^{n+1}}\right|_{1}^{-}. \ n=0,1,\dots \tag{26}$$

Substitution of Eq. (25) into Eq. (26) leads to the general expression

$$\left.\frac{\partial^{n+2}c_{A}}{\partial x^{n+2}}\right|_{1}^{-} = \left.\frac{\partial\left(\partial^{n}c_{A}/\partial x^{n}\right)}{\partial t}\right|_{1} + (p_{e}-v_{s})\left.\frac{\partial^{n+1}c_{A}}{\partial x^{n+1}}\right|_{1}^{-}. \tag{27}$$

The interface concentration of ion $M^+$ and corresponding derivatives up to the third order can be easily written as (from Eqs. (13) and (27)),

$$c_{A}\big|_{1}^{-} = 0, \tag{28}$$

$$\left.\frac{\partial c_{A}}{\partial x}\right|_{1}^{-} = -v_{s}, \tag{29}$$

$$\left.\frac{\partial^{2}c_{A}}{\partial x^{2}}\right|_{1}^{-} = \left.\frac{\partial c_{A}}{\partial t}\right|_{1}^{-} + p_{e}\left.\frac{\partial c_{A}}{\partial x}\right|_{1}^{-} = \left.\frac{\partial c_{A}}{\partial t}\right|_{1} + (p_{e}-v_{s})\left.\frac{\partial c_{A}}{\partial x}\right|_{1}^{-} = -v_{s}(p_{e}-v_{s}), \tag{30}$$

$$\left.\frac{\partial^{3}c_{A}}{\partial x^{3}}\right|_{1}^{-} = \left.\frac{\partial\left(\partial c_{A}/\partial x\right)}{\partial t}\right|_{1} + (p_{e}-v_{s})\left.\frac{\partial^{2}c_{A}}{\partial x^{2}}\right|_{1}^{-} = -\frac{\partial v_{s}}{\partial t} - v_{s}(p_{e}-v_{s})^{2}. \tag{31}$$

In principle, any higher order derivatives ($\left.\frac{\partial^{4}c_{A}}{\partial x^{4}}\right|_{1}^{-}, \left.\frac{\partial^{5}c_{A}}{\partial x^{5}}\right|_{1}^{-}, \dots\dots$) can be obtained in a similar manner. The concentration field can be written in terms of those derivatives through a Taylor expansion,

$$c_A\big|_2^+ = c_\infty = c_A\big|_1^- + \sum_{n=1}^{\infty} \frac{(-L)^n}{n!} \frac{\partial^n c_A}{\partial x^n}\bigg|_1^+ = c_A\big|_1^- - L\frac{\partial c_A}{\partial x}\bigg|_1^+ + \frac{L^2}{2!}\frac{\partial^2 c_A}{\partial x^2}\bigg|_1^+ - \frac{L^3}{3!}\frac{\partial^3 c_A}{\partial x^3}\bigg|_1^+ + \ldots \quad (32)$$

By substituting expressions for interface concentration and derivatives (Eqs. (28)-(31)),

$$c_\infty \approx v_s L - \frac{L^2}{2!} v_s (p_e - v_s) + \frac{L^3}{3!} v_s (p_e - v_s)^2 - \frac{L^4}{4!} v_s (p_e - v_s)^3 + \ldots \quad (33)$$

Sum of the infinite series in Eq. (33) leads to the result

$$c_\infty = \frac{v_s}{p_e - v_s} \{1 - \exp[-(p_e - v_s)L]\}, \quad (34)$$

or equivalently considering the dimensionless parabolic rate constant $k_p = 2v_s L$,

$$c_\infty = \frac{k_p}{2p_e L - k_p} \left\{1 - \exp\left[\frac{k_p}{2} - p_e L\right]\right\}. \quad (35)$$

The algebra of Eq. (34) (or Eq. (35)) can be easily solved in order for a relationship between oxide growth rate $v_s$ (or $k_p$) and the oxide thickness $L$ with known parameters $c_\infty$, $\alpha$, and $L_D$. A plot of the variation of $v_s$, $k_p$, and $p_e$ with oxide thickness $L$ for $c_\infty = 10^{-4}$, $\alpha = L_D = 20$ is shown in Figure 6. The same plot but for a larger Debye-Hückel screening length of $L_D = 200$ is shown in Figure 7. Three stages can be clearly identified:

1) At the early oxidation stage $L < 2L_D$, the péclet number can be approximated by $p_e \approx 2\sinh(\alpha/2L) \gg v_s$ from Eq. (22). Therefore, the oxidation speed $v_s$ can be approximated by

$$v_s \approx c_\infty p_e = 2c_\infty \sinh(\alpha/2L) \quad (36)$$

from Eq. (34). At the very early stage $L < X_1 = \alpha/2$, $v_s \approx c_\infty \exp(\alpha/2L)$ and the inverse logarithm law from Cabrera and Mott's theory is recovered,

$$\log(v_s) = \log(c_\infty) + \alpha/2L, \tag{37}$$

where $X_1$ corresponds to the upper limit of thickness for validity of Cabrera and Mott's theory.

Later for $\alpha/2 < L < 2L_D$, $v_s \approx c_\infty \alpha/L$ and the kinetics can be approximated by a parabolic rate law with the parabolic rate constant being

$$k_p = 2v_s L = 2\alpha c_\infty. \tag{38}$$

This stage is shown in Figs. 6 and 7 where $k_p$ is almost a constant following the inverse logarithm law. In this stage, the kinetic rate law is written as

$$\log(v_s) = \log(\alpha c_\infty) - \log(L), \tag{39}$$

or

$$\log(v_s) = \frac{1}{2}\log\left(\frac{\alpha c_\infty}{2}\right) - \frac{1}{2}\log(t). \tag{40}$$

2) At the intermediate stage where $L > 2L_D$ and $p_e \Box v_s$, the péclet number can be approximated by

$$p_e \approx \frac{\alpha}{2L_D \sinh(L/2L_D)} \approx \frac{\alpha}{L_D}\exp(-L/2L_D) \tag{41}$$

from Eq. (22) and the oxide growth rate can be reduced to

$$v_s \approx c_\infty p_e = \frac{\alpha c_\infty}{L_D}\exp(-L/2L_D). \tag{42}$$

This leads to a direct logarithmic rate law written as

$$\log(v_s) = \log\left(\frac{\alpha c_\infty}{L_D}\right) - \frac{L}{2L_D}, \tag{43}$$

or

$$\log(v_s) = \log(2L_D) - \log(t). \tag{44}$$

In this stage, a decrease of parabolic rate constant $k_p$ with the oxide thickness $L$ is observed. The trend is in agreement with experimental findings [15], though a more extensive and thorough comparison should be in place in our future studies.

3) At the latest stage where $L \gg L_D$ and $p_e \approx 0 < v_s$, Equation (34) can be reduced to

$$c_\infty = \exp(v_s L) - 1. \tag{45}$$

The kinetics is again approximated by a parabolic rate law with the parabolic rate constant being

$$k_p = 2v_s L = 2\log(1 + c_\infty). \tag{46}$$

This stage was also shown in Figures 6 and 7 where $k_p$ is almost a constant but an order of magnitude smaller than the $k_p$ in the first stage (Eq. (38)). The kinetic rate laws are expressed as

$$\log(v_s) = \log(\log(1 + c_\infty)) - \log(L), \tag{47}$$

or

$$\log(v_s) = \frac{1}{2}\log\left(\frac{\log(1+c_\infty)}{2}\right) - \frac{1}{2}\log(t). \tag{48}$$

A typical plot of the variation of oxidation rate $v_s$ with time $t$ for all three stages but after the very early stage is presented in Figure 8, where the initial stage (Eq. (40)), the transitional stage (Eq. (44)), and the final stage (Eq. (48)) are all presented within the same plot. Two lengths that delineate the three stages ($L_1$ and $L_2$ shown in Fig. 8) can be identified by solving kinetic Eqs. (40), (44), and (48) together and the results are

$$L_1 = 2\log(4) L_D \approx 2.77 L_D, \tag{49}$$

and

$$L_2 = 2 L_D \log\left(\frac{4\alpha c_\infty}{\log(1+c_\infty)}\right) \approx 2\log(4\alpha) L_D. \tag{50}$$

Both lengths $L_1$ and $L_2$ are dependent on $L_D$, the Debye-Hückel screening length. $L_2$ is also dependent on $\alpha$.

Finally, the ion concentration profile within the oxide film can be computed from Eq. (34). At the first stage where the péclet number $p_e \gg v_s$, the ionic transport is drift-dominated and the concentration is

$$c_A(x,t) = \frac{v_s}{p_e - v_s}\left\{1 - \exp\left[-(p_e - v_s)x\right]\right\}. \tag{51}$$

At the last stage where the péclet number $p_e \ll v_s$, the ionic transport shifts to diffusion-dominated and the concentration can be written as

$$c_A(x,t) = \frac{v_s}{v_s - p_e}\left\{\exp\left[(v_s - p_e)x\right] - 1\right\}. \tag{52}$$

The concentration profiles are sketched in Figure 1 as the solid line (Eq. (51)) and dashed line (Eq. (52)). A transitional stage occurs for ionic transport shifting from the drift-dominated regime to diffusion-dominated regime.

**IV. Conclusions**

In summary, typical oxidation kinetics exhibits various kinetic rate laws at different oxidation stages. At the very early stage $L < X_1$, the oxidation follows an inverse logarithmic law, where $X_1 = \alpha/2$ is the upper limit of the thickness in Cabrera and Mott's theory. At the final stage $L \gg L_D$, the oxidation follows a parabolic rate law, namely the Wagner's theory. Two lengths $L_1$ and $L_2$ can be identified from the model. In the intermediate stage where $X_1 < L < L_1$, the oxidation follows a parabolic rate law, and when $L_1 < L < L_2$, the oxidation follows a direct logarithmic law that is frequently observed in experiments. The model recovers the classical oxidation models at two extremes and presents the smooth transition between them. The parabolic rate constant $k_p$ is predicted to be decreasing with increasing oxide thickness $L$, also in agreement with experimental findings.

**ACKNOLEDGEMENT**

This research was supported by a grant from the U.S. Department of Energy, Office of Science, Office of Basic Energy Sciences, Materials Science program.

FIG. 1. Schematic plot of metal oxidation due to the ionic transport and the sketch of ion concentration profiles at early stage (solid line) and final stage (dash line).

FIG. 2. Logarithm plot of rate of growth of a hypothetical p-type oxide film as a function of its thickness, calculated using the theory of Cabrera and Mott when thin ($X < X_1$) and of Wagner when thick ($X > L_D$). The parameters used are appropriate to a film of NiO growing by lattice diffusion at 500 ºC. (A. Atkinson, Rev. Mod. .Phys. Vol. 57 pp. 437, Copyright 1985, reproduced or modified by permission of APS).

FIG. 3. Schematic plot of potential energy landscape for ion transport under electrical field $E$.

FIG. 4. Schematic plot of the moving interface used to derive the differential relationships between interface values and interface velocity (Eqs. (23) and (24)).

FIG. 5. Spatial distribution of electrical potential field $\varphi$ for various oxide thickness of $L/L_D = 1, 5, 10$ and $100$ (Eq. (17)).

FIG. 6. Variation of $v_s$, $k_p$, and $p_e$ with the oxide thickness $L$ for given parameters $c_\infty = 10^{-4}$, $\alpha = L_D = 20$. Plot shows the decrease of $k_p$ with thickness $L$.

FIG. 7. Variation of $v_s$, $k_p$, and $p_e$ with the oxide thickness $L$ for given parameters $c_\infty = 10^{-4}$, $\alpha = 20$, $L_D = 200$.

FIG. 8. Plot of variation of $v_s$ with time $t$ exhibits various kinetic rate laws at different oxidation stages. Two lengths $L_1$ and $L_2$ can be identified from the plot to delineate different stages.

Fig. 1

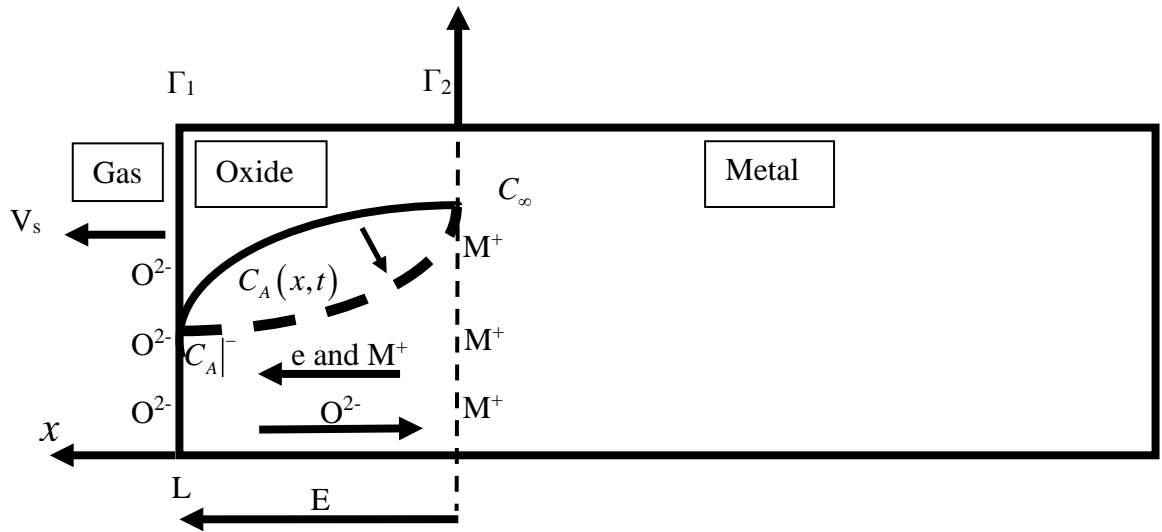

Fig. 2

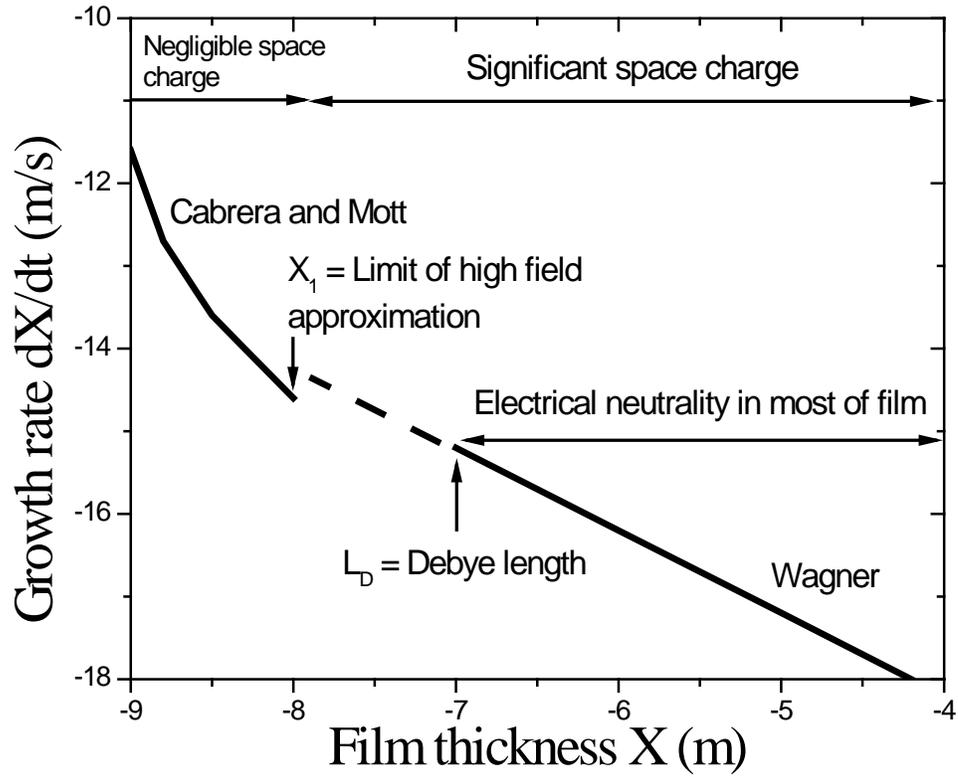

Fig. 3

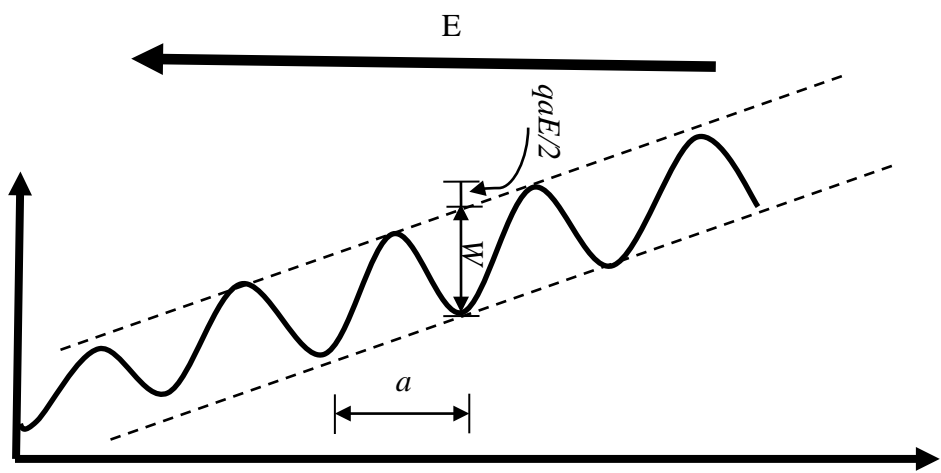

Fig. 4

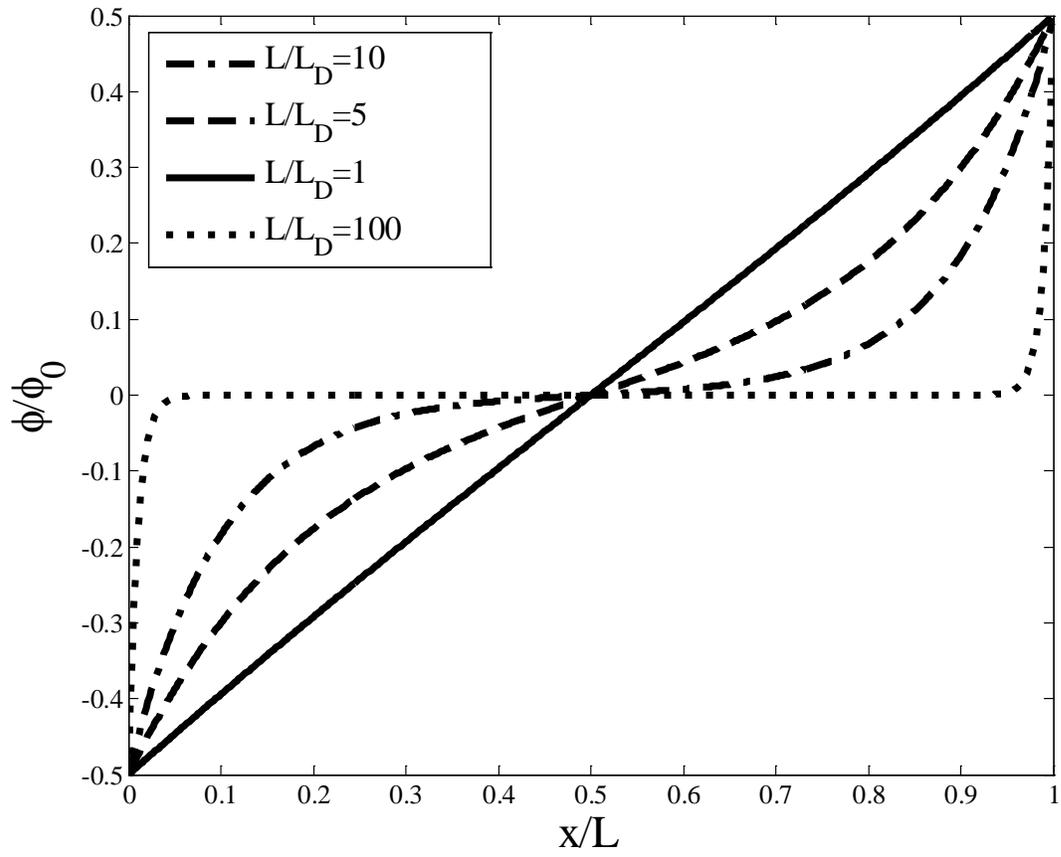

Fig. 5

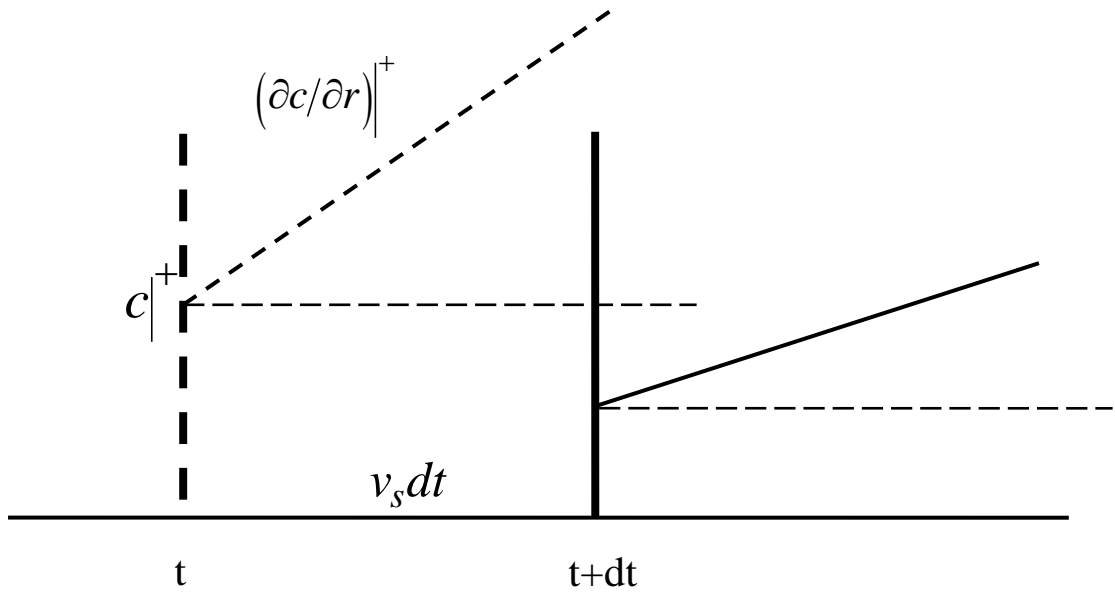

Fig. 6

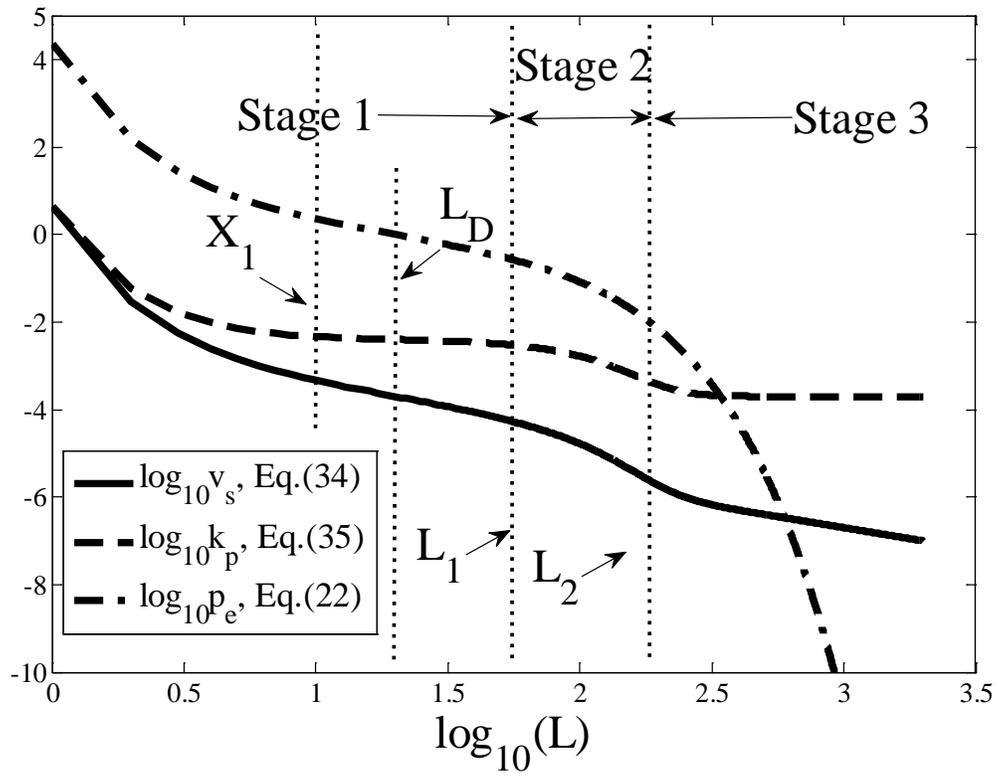

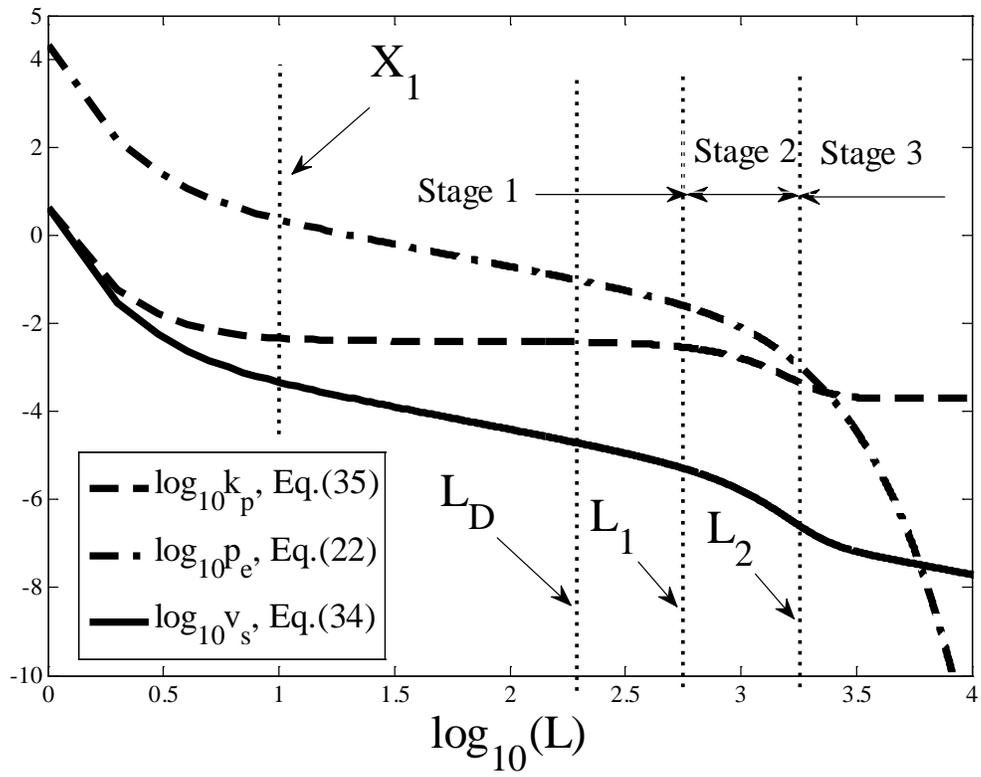

Fig. 8

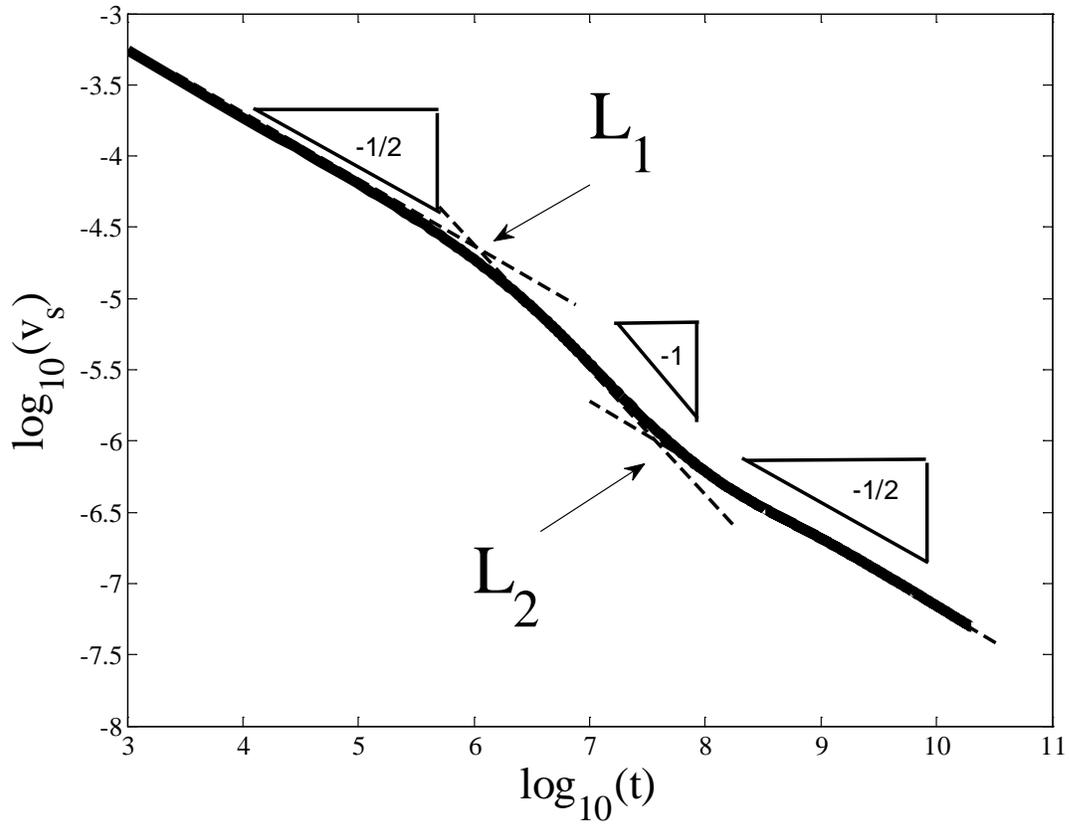

**References**

1. Z. Xu and P. Meakin, *J. Chem. Phys.*, 2008, **129**, 014705.
2. Z. Xu and P. Meakin, *J. Chem. Phys.*, 2011, **134**, 044137.
3. A. Atkinson, *Rev. Mod. Phys.*, 1985, **57**, 437-470.
4. G. Bakradze, L. P. H. Jeurgens, T. Acarturk, U. Starke and E. J. Mittemeijer, *Acta Materialia*, 2011, **59**, 7498-7507.
5. G. Tammann, *Z. Anorg. Allgem. Chem.*, 1920, **III**, 78.
6. N. B. Pilling and R. E. Bedworth, *J. Inst. Metals*, 1923, **29**, 529-582.
7. C. Wagner, *Phys. Chem.*, 1933, **21B**, 25.
8. N. Cabrera and N. F. Mott, *Rep. Prog. Phys.*, 1948, **12**, 163-184.
9. W. Scheuble, *Zeitschrift für Physik A Hadrons and Nuclei* 1953, **135**, 125-140.
10. R. K. Hart, *Proceedings of the Royal Society of London Series a-Mathematical and Physical Sciences*, 1956, **236**, 68-88.
11. J. Kruger and H. T. Yolken, *Corrosion*, 1964, **20**, 29t.
12. B. Lustman and R. F. Mehl, *Transactions of the American Institute of Mining and Metallurgical Engineers*, 1941, **143**, 246-271.
13. A. Hasnaoui, O. Politano, J. M. Salazar and G. Aral, *Phys. Rev. B*, 2006, **73**.
14. K. R. Lawless, *Rep. Prog. Phys.*, 1974, **37**, 231-&.
15. D. Monceau, R. Peraldi and B. Pieraggi, *Oxidation of Metals*, 2002, **58**, 275-295.
16. Z. J. Xu, H. Huang, X. Y. Li and P. Meakin, *Comput. Phys. Commun.*, 2012, **183**, 15-19.
17. Z. Xu, *J. Heat Transfer*, 2012, **134**, 071705.
18. Z. Xu, K. M. Rosso and S. M. Bruemmer, *J. Chem. Phys.*, 2011, **135**, 024108.